\documentclass[aps, reprint,pra,nofootinbib,superscriptaddress,longbibliography]{revtex4-1}
\usepackage{dcolumn}
\usepackage[utf8]{inputenc}
\usepackage{amsmath}
\usepackage{amsfonts}
\usepackage{amssymb}
\usepackage{graphicx}
\usepackage{hyperref}
\usepackage{xcolor}

\newcommand{\ket}[1]{{\left|{#1}\right\rangle}}
\newcommand{\bra}[1]{\ensuremath{\left\langle{#1}\right |}}

\begin{document}

\title{Repulsive gravitational force as a witness of the quantum nature of gravity}

\author{Pablo L. Saldanha}\email{saldanha@fisica.ufmg.br} \affiliation{Departamento de F\'isica, Universidade Federal de Minas Gerais, Belo Horizonte, MG 31270-901, Brazil}
\author{Chiara Marletto}\affiliation{Clarendon Laboratory, University of Oxford, Parks Road, Oxford OX1 3PU, United Kingdom} \author{Vlatko Vedral}
\affiliation{Clarendon Laboratory, University of Oxford, Parks Road, Oxford OX1 3PU, United Kingdom}

\date{\today}

\begin{abstract}
We show that a single spatially superposed `source' mass acting on a `probe' matter wavepacket can reveal the quantum nature of the gravitational field. For this we use a specific state preparation and measurement of the superposed source mass, including a postselection, which altogether results in a repulsive gravitational force on the probe particle. A classical gravitational field can never lead to repulsion, as the effect requires quantum interference of two distinct states of gravity. The eventual observation of such an effect could be seen as a violation of Einstein's theory of general relativity, where gravity is always attractive. We also present a calculation in the Heisenberg picture under the formalism of weak values that illustrates how repulsion is achieved. Finally, we estimate the range of parameters (masses and the spatio-temporal extent of interference) for which the experiment is feasible. 
\end{abstract}


\maketitle

\section{Introduction}

Quantum theory and general relativity are about one century old and they are successful in their respective domains. Yet, they rely on incompatible principles. Quantum theory treats nature as made up of physical observables that obey the Heisenberg uncertainty principle, so they cannot all be measured simultaneously to arbitrarily high accuracy. In contrast, general relativity describes gravity as a classical entity, where there is no such limitation. This tension signals that our current theories are incomplete and motivates the search for a unified description in which spacetime and quantum phenomena are part of a single coherent framework, as in current quantum gravity proposals \cite{kiefer}. One difficulty with these proposals is that they are extremely difficult to test. Until recently, it was thought that laboratory scale experiments could not probe quantum effects in gravity.

This has been changed by a novel proposal to test quantum effects in gravity via the so-called gravitationally induced entanglement (GIE) effect \cite{marletto17,bose17}. The GIE links gravity with one of the most characteristically quantum phenomena: entanglement. The core idea is that if two massive particles are each placed in a quantum superposition of positions, and if they interact only through their mutual gravitational attraction, then gravity itself can generate entanglement between them. Because entanglement cannot be produced by purely classical interactions, observing this effect would strongly suggest that gravity must possess intrinsically quantum degrees of freedom. What makes the GIE effect especially compelling is that it avoids the need for extreme energies or Planck-scale experiments. Instead, it relies on tabletop-scale setups involving mesoscopic masses, interferometry, and precise control of quantum states, with different variations \cite{marletto17,bose17,marletto18,krisnanda20,tilly21,schut22,pedernales22,feng22,vicentini24,dipietra24,bose25,marletto25}.

Here we propose an interesting variant of the existing proposals to witness quantum gravity effects via GIE, by using the fact that the superposition of a positive force with a null force on a quantum particle may result in a negative momentum transfer to the particle when the appropriate post-selection is made \cite{correa18}. This `quantum interference of force' effect was experimentally verified with photons \cite{militani25} and could result in an effective electrostatic attraction between electric charges of the same sign \cite{cenni19}. As we shall explain, our proposal requires only one mass to be prepared in a superposition, which makes it easier compared to the previous ones, which require two massive superpositions \cite{marletto17,bose17,marletto18,krisnanda20,tilly21,schut22,pedernales22,feng22,vicentini24,dipietra24,bose25,marletto25}.

In our proposed scheme, a `source' massive quantum particle is put in a superposition of two different spatial locations. If gravity obeys the quantum superposition principle, a `probe' massive quantum particle will be subject to the superposition of two attractive gravitational forces, each associated with a possible location of the source particle. We show that, with the appropriate post-selection of the quantum state of the source particle, the probe will experience an effective gravitational repulsion. In this case, if the experiment is repeated with many particles, there will be a momentum transfer to the probe particles in the opposite direction of the gravitational force in the ensemble -- a behaviour with no classical analogue. This anomalous momentum transfer would be a direct consequence of the entanglement generated between the source particle and the probe particle mediated by the gravitational field, which is a sufficient condition for witnessing the quantum nature of the gravitational field \cite{marletto17,bose17}. But in our scheme it is not necessary to measure quantum correlations between the particles, which is also a great simplification in relation to the previous proposals \cite{marletto17,bose17,marletto18,krisnanda20,tilly21,schut22,pedernales22,feng22,vicentini24,dipietra24,bose25,marletto25}.

\section{Proposal to observe a repulsive gravitational force}

Consider the scheme depicted in Fig. 1. We have a Mach-Zehnder interferometer for the source quantum particle with mass $M$, with the probe quantum particle with mass $m$ alongside it. We could have both particles in a free fall situation, with the scheme of Fig. 1 showing their trajectories, or in a trap that keeps their $y$ and $z$ coordinates fixed, with time being represented in the vertical down direction in the figure. At time $t_1=0$, the source particle is in a superposition state  $\alpha\ket{A}+\beta\ket{B}$, with $\ket{A}$ being a state localized around a position $x_A$ and $\ket{B}$ a state localized around a position $x_B$. The state of the probe particle is written in terms of the $x$-component of its momentum wavefunction $\psi(p)$, with position wavefunction centered at $x=0$, such that at time $t_1=0$ the system quantum state is
\begin{equation}\label{psi1}
	\ket{\Psi(0)}=\big[\alpha\ket{A}+\beta\ket{B}\big]\otimes\int dp\, \psi(p)\ket{p}.
\end{equation}

\begin{figure}
    \centering
    \includegraphics[width=9cm]{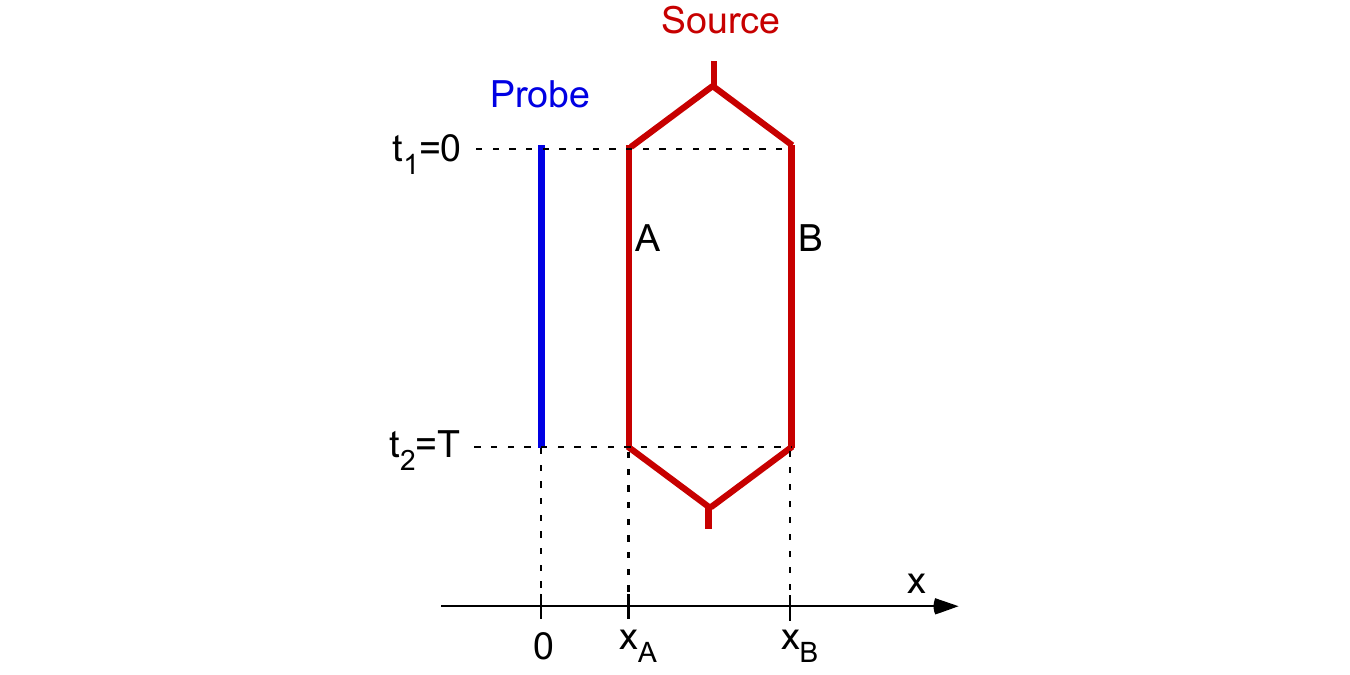}
    \caption{Scheme to test a quantum behaviour of gravity using the quantum interference of force effect \cite{correa18,cenni19}. The source quantum massive particle is put in a superposition of two different spatial locations centered in $x_A$ and $x_B$. The quantum superposition of the two possible gravitational attractions in a probe quantum particle, relative to the two possible positions of the source particle, can result in an effective gravitational repulsion, depending on the post-selection of the source particle state. This is a behaviour with no classical analogue.}
 \end{figure}

Between times $t_1=0$ and $t_2=T$, the quantum particles gravitationally attract each other. Let us see the consequences of the assumption that gravitational forces obey the quantum superposition principle.  Consider that the source particle has a quantum state with a very large momentum uncertainty, such that the change on its quantum state due to this interaction is negligible. The momentum wave function of the probe particle, on the other hand, may be affected by this interaction. Hence the system quantum state at time  $t_2=T$ can be written as
\begin{equation}\label{psi2}
	\ket{\Psi(T)}=\int dp\big[\alpha e^{i\phi_A}\psi(p-\delta_A)\ket{A}+\beta e^{i\phi_B}\psi(p-\delta_B)\ket{B}\big]\ket{p}
\end{equation}
with
\begin{equation}\label{delta}
	\delta_j=\frac{GMmT}{x_j^2},
\end{equation}
where $G$ is the gravitational constant and $j=\{A,B\}$, assuming that $x_A$, $x_B$, and $x_B-x_A$ are much larger than the width of the position wave functions of the quantum particles. Here $\phi_j$ are phases that come from the interaction. We see that the probe particle feels the superposition of two momentum transfers of different magnitudes in the positive $x$ direction, which results in entanglement between the source and the probe particles mediated by the gravitational field.

Suppose now that the source particle is post-selected on the state $[-\ket{A}e^{i\phi_A}+\ket{B}e^{i\phi_B}]/\sqrt{2}$ at the exit of the Mach-Zehnder interferometer. The momentum wave function of the probe particle in this branch is 
\begin{equation}\label{psips}
	\psi_{p.s.}(p)\propto\beta\psi(p-\delta_B)-\alpha\psi(p-\delta_A).
\end{equation}
If $\alpha$ and $\beta$ are both real and positive, with $\beta>\alpha$, we may have a resulting momentum wave function with negative average momentum, as depicted in Fig. 2. In this case, the superposition of two gravitational attractions of different magnitude may result in an effective gravitational repulsion. As is evident from Fig. 2, this is a simple quantum interference effect, with the wavefunction $\alpha\psi(p-\delta_A)$ subtracting more positive momenta than negative momenta from the wave function  $\beta\psi(p-\delta_B)$ through destructive interference \cite{correa18}. 

\begin{figure}
    \centering
    \includegraphics[width=9cm]{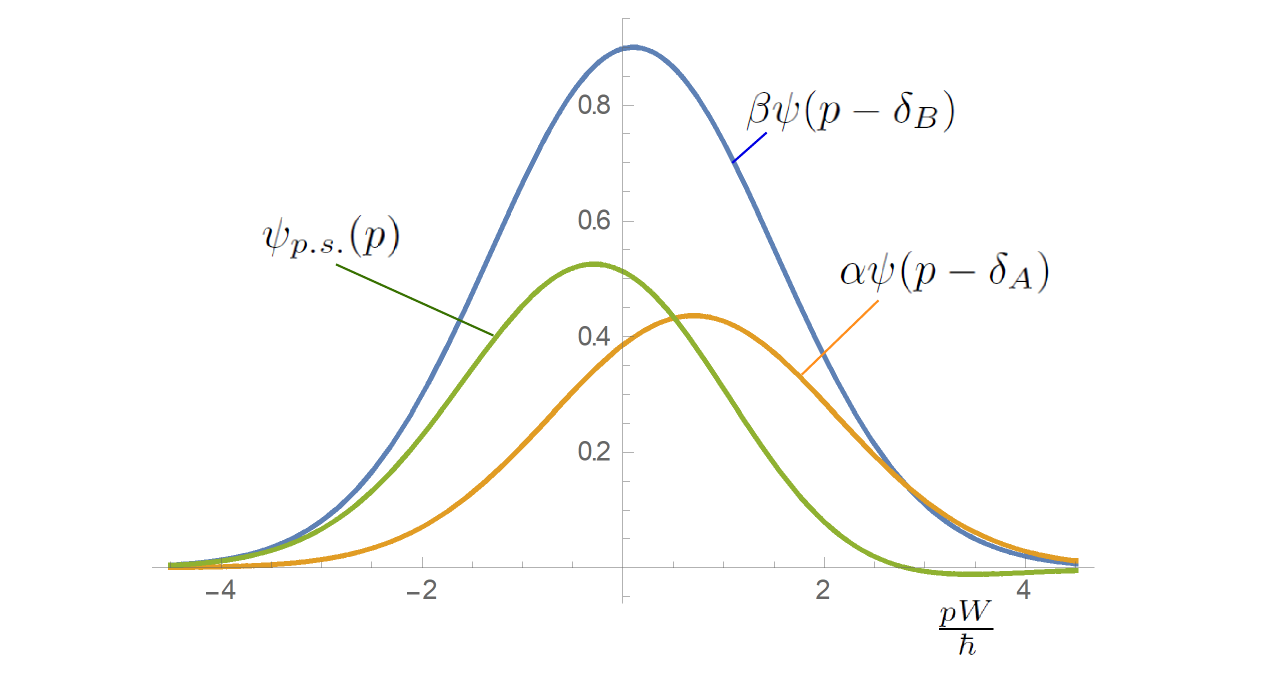}
    \caption{Decomposition of the post-selected wavefunction of the probe particle from Eq. (\ref{psips}), showing how the superposition of two wavefunctions with average positive momenta can generate a wavefunction with negative average momentum through destructive interference. We considered $\psi(p)\propto\mathrm{Exp}[-p^2W^2/(4\hbar^2)]$, $\beta=0.9$, $\alpha=\sqrt{1-\beta^2}$, $\delta_B=0.1\hbar/W$, and $\delta_A=0.7\hbar/W$.}
 \end{figure}

Since the gravitational attraction between two quantum particles is expected to be very small, we can make a series expansion of $\psi_{p.s.}(p)$ from Eq. (\ref{psips}) around $p-\delta_B$ to write
\begin{eqnarray}\label{psipsweak}\nonumber
	&&\psi_{p.s.}(p)\propto(\beta-\alpha)\left[\psi(p-\delta_B)+\frac{\alpha}{(\beta-\alpha)}(\delta_A-\delta_B)\frac{d\psi}{dp}\right]\\
	&&\propto\psi(p-\delta_{ef})\,,\;\mathrm{with}\;\;\delta_{ef}=\delta_B-\frac{\alpha}{(\beta-\alpha)}(\delta_A-\delta_B).
\end{eqnarray}
$\delta_{ef}$ is the effective momentum transfer, which is in the opposite direction of the applied force for $\alpha(\delta_A-\delta_B)/(\beta-\alpha)>\delta_B$. We can also write the effective momentum transfer as $\delta_{ef}=\delta_B-  (\delta_A-\delta_B)\langle \hat{\Pi}_A\rangle_W$, where
\begin{equation}\label{weak}
	\langle \hat{\Pi}_A\rangle_W = \frac{\bra{\Psi_f}\hat{\Pi}_A{\ket{\Psi_i}}}{\langle\Psi_f|\Psi_i\rangle}=-\frac{\alpha}{(\beta-\alpha)}
\end{equation}
is the weak value of the projector $\hat{\Pi}_A\equiv|A\rangle\langle A|$ with the pre-selected state $\ket{\Psi_i}=\alpha\ket{A}+\beta\ket{B}$ and the post-selected state $\ket{\Psi_f}=[-\ket{A}+\ket{B}]/\sqrt{2}$ for the source quantum particle \cite{aharonov88,dressel14}. Note that if $\ket{\Psi_i}$ and $\ket{\Psi_f}$ are almost orthogonal states, $\langle A\rangle_W$ from Eq. (\ref{weak}) can be large and the momentum displacement on the wavefunction of Eq. (\ref{psipsweak}) can be orders of magnitude higher than the individual momentum transfers $\delta_A$ and $\delta_B$, being also in the opposite direction of the applied gravitational forces. For instance, for $\beta=1/\sqrt{2}+0.007$, $\alpha=\sqrt{1-\beta^2}$, and $\delta_A=10\delta_B$, the effective momentum transfer to the probe particle according to Eq. (\ref{psipsweak})  is around $-50\delta_A$. So, the anomalous momentum transfer can be amplified by weak value amplification \cite{dressel14}, which can facilitate its experimental verification. An inconvenience is that the higher the weak value amplification, the lower the probability of performing the desired post-selection. For the parameters we considered, the probability of the post-selection is $|\langle\Psi_f|\Psi_i\rangle|^2\approx10^{-4}$. Since the increase of the conditional shift due to weak value amplification is accompanied by a reduced number of successful post-selections, it generates no gain in the system statistical sensitivity \cite{ferrie14}. But depending on the experimental scheme, the weak value amplification may allow the measurements of quantities that would be difficult  (or even impossible) to observe with other schemes \cite{dressel14}.

The effect can also be explained in the Heisenberg picture. We only do this in order to shed further light on the effect, but the conclusions will be the same as in the analysis above. We start with the system Hamiltonian, given by
\begin{equation}
\hat{H} = \frac{\hat{P}}{2M} + \frac{\hat{p}^2}{2m} - \frac{GmM}{|\hat{X}-\hat{x}|} \; ,
\end{equation}
where $\hat{X}$ and $\hat{P}$ are the position and momentum operators for the source particle, with mass $M$, and $\hat{x}$ and $\hat{p}$ are the position and momentum operators for the probe particle, with mass $m$. The first two terms are the kinetic energies of the source and probe particles, while the last term designates their gravitational interaction. We are interested in the Heisenberg equation of motion for the probe momentum, which is
\begin{equation}
\dot {\hat{p}} =\frac{i}{\hbar}[\hat{H},\hat{p}]= \frac{GmM(\hat{X}-\hat{x})}{|\hat{X}-\hat{x}|^3} \; .
\end{equation}
Considering the physical situation depicted in Fig. 1, where the probe particle is in a quantum state localized around $x=0$ and the source particle is in a superposition state with wavefunctions localized around $X=x_A$ (state $\ket{A}$) and $X=x_B$ (state $\ket{B}$), the momentum transfer to the probe particle after an interaction time $T$ can be written as
\begin{equation}
\Delta \hat{p}=\int_0^T\dot {\hat{p}}\, dt\approx \delta_A |A\rangle\langle A| +  \delta_B |B\rangle\langle B|\; ,
\end{equation}
with $\delta_j$ given by Eq. (\ref{delta}). 
If, as before, the source starts in the state $\ket{\Psi_i}=\alpha\ket{A}+\beta\ket{B}$ and after the interaction the state $\ket{\Psi_f}=[-\ket{A}+\ket{B}]/\sqrt{2}$ is postselected, the weak value of the momentum transfer $\Delta \hat{p}$ is
\begin{equation}
	\langle \Delta \hat{p}\rangle_W = \frac{\bra{\Psi_f}(\Delta \hat{p}){\ket{\Psi_i}}}{\langle\Psi_f|\Psi_i\rangle}=\frac{\beta\delta_B-\alpha\delta_A}{(\beta-\alpha)},
\end{equation}
identical to momentum transfer $\delta_{ef}$ from Eq. (\ref{psipsweak}). It is worth mentioning that the use of the weak value formalism to compute the momentum transfer to the probe particle is only possible when the involved momentum transfers are much smaller than the particle initial momentum uncertainty \cite{aharonov88,dressel14}, which was also considered to obtain Eq. (\ref{psipsweak}). This concludes our analysis in the Heisenberg picture. 

\section{Experimental feasibility of the proposal}

We now discuss the feasibility of this experimental proposal. Recent works are based on the possibility of performing a Mach-Zehnder interferometer with quantum particles with masses of the order of $10^{-14}$kg \cite{bose17,vicentini24,dipietra24}, which could be used as the source particles in the scheme of Fig. 1. This source particle could be a nanodiamond with a single nitrogen-vacancy (NV) centre. Inhomogenous magnetic fields could produce forces on the NV-centre spin, dislocating the nanodiamond in a way that depends on its spin state \cite{wan16,bose17}. In this way, with the spin-dependent forces acting on the NV-centre spin prepared in a suitable initial state, the quantum state of Eq. (\ref{psi1}) could be prepared with the nanodiamond being the source particle, together with an extra probe particle. In principle, the distance $x_B-x_A$ in Fig. 1 could be of the order of hundreds of micrometers \cite{bose17,vicentini24,dipietra24}. 

The probe particle in Fig. 1 could be an atom or we could use a Bose-Einstein condensate, with a set of atoms with identical wave functions, to perform the experiment with an ensemble of probe particles at once. Let us consider cesium atoms, mass $m=2.3\times10^{-25}$kg, with a Gaussian wavefunction $\phi(x)\propto \mathrm{Exp}[-x^2/W^2]$ in position space, corresponding to a wavefunction $\psi(p)\propto\mathrm{Exp}[-p^2W^2/(4\hbar^2)]$ in momentum space, as the probe quantum particles. The initial momentum uncertainty of the probe particles is $\Delta p =\hbar/W$ in this case, while the initial position uncertainty is $\Delta x=W/2$. If we write the effective anomalous momentum transfer of Eq. (\ref{psipsweak}) as $\delta_\mathrm{ef}=-g\delta_A$, with $g$ representing the weak value amplification factor, using Eq. (\ref{delta}) the ratio between the effective momentum transfer of Eq. (\ref{psipsweak}) and the initial momentum uncertainty becomes
\begin{equation}\label{ratio}
	\frac{\delta_\mathrm{ef}}{\Delta p}=-\frac{gGMmWT}{\hbar x_A^2}.
\end{equation}

If the probe atoms are released from a trap at time $t_1=0$, their wavefunction starts to spread in position space, with a width given by $\Delta x(t)=\sqrt{1+4\hbar^2t^2/(m^2W^4)}W/2$ \cite{cohen}, increasing by a factor $\sqrt{2}$ at a time $\tau=mW^2/(2\hbar)$. For $W=10\mu$m, we have $\tau\approx0.1$s for cesium atoms. Let us set $T=\tau$ in Eq. (\ref{ratio}) for our estimation (a time not so large for the wave function to spread much and not so short for the interaction to be completely negligible). After a time considerably larger than $\tau$, the position of the quantum particles would be associated to their velocity (and momentum) at the initial time, such that the momentum distribution of the atoms could be measured, as is usual with Bose-Einstein condensates experiments. For instance, for $t=20\tau=2$s, we have $\Delta x\approx 100\mu$m for the considered parameters and the atomic cloud density can be visualized by shinning light on it, being proportional to the scattered light intensity at each position. The light intensity scattered at each $x$ position will then be related to the presence of a particular momentum component $p$ of the ensemble, with $p\approx mx/t$. Considering that a ratio $\delta_\mathrm{ef}/\Delta p\approx10^{-3}$ is measurable, for $x_A=50\mu$m and $g=100$, according to Eq. (\ref{ratio}) we would need $M\approx2\times10^{-7}$kg, a mass seven orders of magnitude larger than the one from the nanodiamond proposals \cite{bose17,vicentini24,dipietra24}.

One possibility to overcome this situation is to use a probe quantum object with a larger mass. This is considerably easier because the probe does not need to be in a spatial superposition. For instance, if we use the parameters $W=0.5\mu$m, $T=1.2$s,  $x_A=2\mu$m, $g=100$, and consider the mass of the source particle $M=10^{-14}$kg and the mass of the probe particle $m=10^{-20}$kg in Eq. (\ref{ratio}), we obtain $\delta_\mathrm{ef}/\Delta p\approx10^{-3}$. We have to wait a longer time for a probe particle with a larger mass to expand, of the order of 100 seconds or more for the proposed parameters, since we have $\tau\approx 12$s in this case. 


Our preliminary parameter analysis is encouraging. However, just like in the case of the original proposals, one must conduct an extended feasibility study, e.g. to differentiate our effect from other competing effects, such as the Casimir-Polder interaction \cite{bose25,marletto25,casimir48}. The parameters needed to observe a repulsive gravitational force as proposed here are similar to the ones needed to observe the gravitationally induced entanglement (GIE) effect \cite{marletto17,bose17}. The gravitational repulsion can be observed if the ratio $\delta_\mathrm{ef}/\Delta p$ from Eq. (\ref{ratio}) is within the experimental resolution. In the GIE schemes, the effect of the gravitationally induced phase $\phi=GMmT/(\hbar x_A)$ between particles of masses $M$ and $m$ separated by a distance $x_A$ interacting during a time $T$ in two interferometers must generate correlations between the particles that can be measured within the experimental resolution. The difference between these two cases, which we may quantify by the ratio $(\delta_\mathrm{ef}/\Delta p)/\phi$, is a factor $gW/x_A$, considering identical parameters in both cases. In the GIE proposals $W$ must be very small, since one needs to have a large momentum uncertainty such that the gravitational interaction induces the phase $\phi$ without changing the momentum distribution of each particle, since this influence would decrease the entanglement between the particle paths. But in the current proposal, it is not necessary that $W$ be much smaller than $x_A$. It must be smaller, so that the wave functions of the source and probe particles do not overlap, but not much smaller. We assumed that $W\ll x_A$ in the calculations to arrive at analytical results. But as shown in the Appendix with a numerical calculation, the effect persists with $x_A=3W$. The weak value amplification factor $g$, on the other hand, can be large. But similar weak value amplification schemes can also be used in GIE schemes \cite{feng22}, so that similar parameters can be used in both cases. The main technical advantages of our proposal to observe a quantum behavior of gravity in relation to the previous ones \cite{marletto17,bose17,marletto18,krisnanda20,tilly21,schut22,pedernales22,feng22,vicentini24,dipietra24,bose25,marletto25} are: 1) only one mass needs to be in superposition, not two, and 2) there is no need to measure quantum correlations between the quantum particles, only the momentum distribution of the probe particle at the end. We believe that these two advantages increase the feasibility of such an experiment.

\section{Discussion}

Gravity is always attractive, both in Newton's and Einstein's theories. In our proposal, depicted in Fig. 1, a procedure performed in one region of space, consisting in putting a massive source particle in a superposition of two different positions and post-selecting its desired final state, can generate an effective gravitational repulsion in a probe particle localized in another region of space. The eventual observation of a repulsive gravitational force as proposed here could be seen as a violation of Einstein's theory of general relativity.  But it is true that gravitational repulsion is achieved probabilistically. If both possible final states of the source particle are taken into account, the average gravitational force is attractive.  Even so, no classical treatment of gravity can explain this behavior. 

The original GIE proposals’ rely on a well-defined no-go structure: under the assumptions of initially separable probes, no non-gravitational interaction channel, and a local classical mediator, entanglement cannot be generated \cite{marletto17,bose17}. So, the observation of gravitationally induced entanglement in the proposed scheme would attest a quantum behavior of the mediator, in the sense that the gravitational interaction of a massive particle in a quantum superposition of different positions would also be in a quantum superposition. In the scheme presented here, we have a similar logical argument. If the gravitational interaction cannot be in a quantum superposition, a massive particle could only generate a statistical mixture of gravitational attractions in a probe particle. But a statistical mixture of gravitational attractions can only generate a gravitational attraction. To obtain a repulsion, even probabilistically, there must be a quantum superposition of gravitational interactions generating a superposition of momentum transfers to the probe particle, as depicted in Fig. 2. So, the eventual observation of an effective repulsive gravitational force would be a witness of the quantum nature of gravity, in the same way as the original GIE proposals \cite{marletto17,bose17}.

It is clear that quantum technologies are getting closer to being able to test the eventual quantum nature of gravity and related ideas, and it seems to us that it is only a matter of time before the right platform will enable us to answer this fundamental question. We hope and expect that the present proposal brings us closer to witnessing quantum effects in gravity. 

PLS was supported by the Brazilian agency CNPq (Conselho Nacional de Desenvolvimento Científico e Tecnológico).

\appendix*
\section{Numerical calculation of the repulsive gravitational force}

Here we present a numerical calculation showing that the condition that the probe quantum particle wavefunction width $W$ needs to be much smaller than $x_a$, as assumed to arrive at the effective momentum transfer $\delta_{ef}$ from Eq. (\ref{psipsweak}), is not necessary for a repulsive gravitational force to appear upon post-selection. $W$ must be smaller than $x_a$, so that the wave functions of the source and probe particles do not overlap, but not much smaller. Consider an initial Gaussian wavefunction $\phi(x)\propto \mathrm{Exp}[-x^2/W^2]$ for the probe particle  in the scheme of Fig. 1. We assume a situation where the interaction time $T$ between the source and probe particles is much smaller than the time $\tau$ the wave function $\phi(x)$ takes to spread appreciably. Let us also consider that $x_b\gg x_a$, such that if the source particle propagates through path $b$, the probe wavefunction is not changed.  If the source particle propagates through path $a$, due to the interaction energy $GMm/(x-x_a)$ acting for a time $T$, where $M$ and $m$ are the masses of the source and probe particles, the wavefunction modifies to
\begin{equation}
    \phi_a(x)\propto\mathrm{Exp}\left[-\frac{x^2}{W^2}\right]\mathrm{Exp}\left[\frac{-iGMmT}{\hbar(x-x_a)}\right].
\end{equation}
Using the parameters $M=10^{-14}$kg, $m=10^{-20}$kg, $W=0.5\mu$m,   $x_A=3W=1.5\mu$m, and $T=1$s, a numerical calculation of the expectation value of the momentum of the wavefunction $\phi_a(x)$, given by
\begin{equation}
    \delta_a=\int \phi_a^*(x)\frac{\hbar}{i}\frac{d\phi_a(x)}{dx}dx,
\end{equation}
results in $\delta_a=1.6\times10^{-5}\Delta p$, where $\Delta p=\hbar/W$ is the momentum uncertainty of the wavefunction $\phi(x)$. Considering the post-selected wavefunction $\phi_{p.s.}(x)=\beta\phi(x)-\sqrt{1-\beta^2}\phi_a(x)$ with $\beta=1/\sqrt{2}+0.0035$, according to Eq. (\ref{psipsweak}) we would have $\delta_{ef}=-100\,\delta_a$, while a numerical calculation for the momentum expectation value results in $\delta_{ef}'=-104\,\delta_a$. We thus see that the repulsive gravitational force appears with $x_a$ and $W$ being of the same order of magnitude.


%

\end{document}